\documentclass{MechatronikTagung}
\usepackage[utf8]{inputenc} 
\usepackage[T1]{fontenc}

\usepackage{amsmath} 
\usepackage{amssymb} 

\usepackage{graphicx}

\usepackage{booktabs} 
\usepackage{cite} 

\titeldeutsch{Predictive Energy Management for Recuperation Axles in Refrigerated
Trailers
}

\autorenliste{%
Dennis Bank$^1$,
Simon F.G. Ehlers$^1$, 
Karl-Philipp Kortmann$^1$,
Tobias Zeller$^2$,
Patrick Cujic$^2$,
and Thomas Seel$^1$ 
\vspace{2mm}\\
$^1$ Institute of Mechatronic Systems, Leibniz University Hannover, 30823 Garbsen, Deutschland;\\
$^2$ Institute of Electrical Engineering, Karlsruhe
Institute of Technology, 76131 Karlsruhe, Germany
 \\ Contact: dennis.bank@imes.uni-hannover.de
}

\kurzfassungenglisch{%
Refrigerated truck trailers are currently mainly operated with environmentally harmful diesel units; an alternative is to operate the refrigeration unit with electrical energy. However, this requires a battery, the size of which can be reduced by using a recuperation axle, which recovers energy during braking. Current systems work purely reactively and often in so-called towing mode, in which a generator torque is provided without a braking request from the driver in order to secure the energy supply. However, this drag leads to additional consumption in the truck. This work quantifies the potential of predictive energy management that uses route and environmental data to minimize $\mathrm{CO}_2$ emissions. This was done using simulation data obtained with the help of VECTO. It was shown that there is still considerable potential for savings, so this paper provides an important basis for the later development of predictive energy management and, thus, for the electrification of refrigerated truck transports.
}

\begin{document}

\section{Introduction}
Refrigerated transport is crucial for perishable goods like food, pharmaceuticals, and other temperature-sensitive items. Most refrigerated trailers use diesel generators. However, this has considerable environmental implications, including the emission of substantial amounts of $\mathrm{CO}_2$ and $\mathrm{NO}_2$ and contributing to noise pollution \cite{co2}. 

The legal regulations that compel vehicle manufacturers to produce more efficient vehicles do not currently extend to the manufacturers of trailers for commercial vehicles. This is particularly important given the environmental impact of these trailers. As a result, there is a growing interest in developing more environmentally friendly and quieter alternatives to these diesel-powered refrigerated truck trailers.

The challenge lies in simply replacing the Diesel tank and generator with a large battery to power the Trailer Refrigeration Unit (TRU), which is not a viable solution. The size and weight of such a battery would significantly reduce the loading capacity of the trailer, rendering the system economically unfeasible. \\

\begin{figure}[]
\centering

\includegraphics[width=0.45\textwidth]{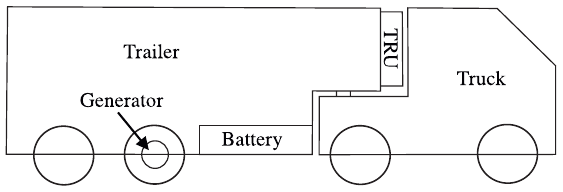}
\vspace{-3mm}
\caption{Overview of an electrified, refrigerated trailer. A generator is added in one of the axles to reduce the battery size and ensure the supply of energy to the TRU during the whole drive}
\vspace{-5mm}
\label{overview}

\end{figure}

In response to this challenge, current systems are exploring the addition of a generator into an axle of the trailer, as shown in Figure \ref{overview}. These systems can recuperate energy during braking as well as generate energy during driving through a process called towing mode. In this mode, the axle generates torque without a braking request, leading to additional energy consumption in the truck but ensuring sufficient energy is always available.

Supply security is a crucial factor for adopting any system, especially those that risk battery depletion, which could interrupt the cooling chain and potentially result in the economic loss of the entire load. \\Therefore, a reliable and constant energy supply is vital in designing and implementing these solutions. 

Nowadays, the Energy Management Systems (EMS) used in these trailers operate reactively, without any foresight of the upcoming drive, the potential energy that is recuperated, or the duration of driving breaks during which no energy can be generated. These systems typically control the towing mode by activating and deactivating it based on two limits for the battery's State of Charge (SoC). They also turn off the towing mode during uphill drives or at specific speeds when the truck requires full power. While many works have been published showing the potential for predictive EMS in multi-energy storage vehicles \cite{wu}, such as hybrid vehicles have been published for cars, to the best of our knowledge, there is no previous work showing the benefits for an electrified truck trailer. To ensure that the system has sufficient energy for the entire drive, the towing mode is used excessively, resulting in higher than necessary fuel consumption in the truck.

\begin{figure}[]
\centering

\includegraphics[width=0.45\textwidth]{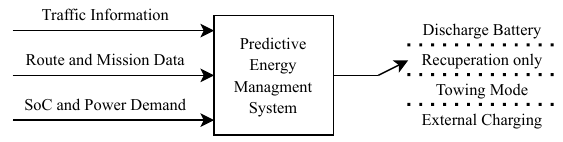}
\vspace{-3mm}

\caption{Schematic representation of a predictive EMS. It could use different information about the route, current traffic, etc., to optimize the selection of the current operating mode and thus reduce the emissions of the system}
\label{figure1}
\vspace{-6mm}
\end{figure}

This illustrates the need for a predictive EMS, of which a schematic is shown in Figure \ref{figure1}. It reduces the amount of towing during a drive cycle and ensures that it only happens in the most efficient route sections to minimize the additional energy consumption induced in the truck, thus reducing the overall emissions of the system as well as the operating costs. The EMS could take many different sources of data into account, such as current traffic and route, as well as mission-specific information such as an estimated stopping time of the truck for loading and unloading, as well as information about the amount of load and the current weather to estimate the power demand of the TRU over the drive. In order to demonstrate the potential savings caused by a predictive EMS, this paper uses drive cycles and vehicle models from the Vehicle Energy Consumption Calculation TOol (VECTO) \cite{vecto}.

\section{Trailer \& Drivecycle Modelling}
This section briefly introduces VECTO, the used drive cycles, and the modifications made to expand the simulation to recuperating trailers.

\subsection{VECTO}
VECTO is a simulation software developed by the European Commission. \\ Its primary function is to determine $\mathrm{CO}_2$ emissions and fuel consumption from Heavy Duty Vehicles (HDVs), which include trucks, buses, and coaches with a gross vehicle weight above 3.5 tons \cite{vecto}. \\

VECTO uses various vehicle information, including factors such as rolling resistance, air drag, transmission data, and an engine map. It then simulates $\mathrm{CO}_2$ emissions and fuel consumption based on vehicle longitudinal dynamics. This is achieved using a generic driver model for the simulation of different drive cycles, which vary depending on the specific operation scenarios of the vehicle.

One of the critical features of VECTO is its inclusion of different example vehicle definitions for different vehicle and truck types. Furthermore, VECTO includes multiple drive cycles for different operation scenarios, enhancing its versatility.

The simulated fuel consumption of VECTO has been validated with real-world tests, demonstrating its accuracy and reliability. The measured fuel consumption deviated less than 5\% from the simulated values \cite{vecto}, making VECTO an effective tool for the fuel consumption simulation of HDVs. It was thus chosen as the data basis for this paper.






\subsection{Drive Cycles}

Three different drive cycles from VECTO were used to evaluate the differences between additional fuel consumption to represent the multitude of different driving scenarios. The urban delivery, regional delivery, and long haul reference cycles were used \cite{vecto}. As they do not take up a whole working day, they were repeated multiple times, interrupted by a 45min break between repetitions in order to simulate loading and unloading breaks, as stops taking between thirty minutes and one hour are the most frequent stops in all drive cycles \cite{kit_stop}. 

\begin{figure}[h]
\centering

\includegraphics[width=0.45\textwidth]{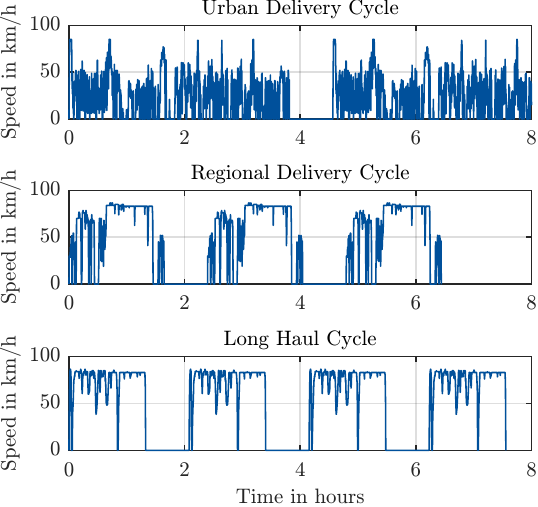}
\vspace{-3mm}

\caption{The three modified drive cycles represent urban, regional, and long-haul traffic deliveries. As can be seen, the average speed and the driving dynamics vary widely between the different drive cycles, giving good coverage of possible driving scenarios.}
\label{cycles}

\end{figure}

The three cycles are shown in Figure \ref{cycles}. As can be seen, the average speed and driving dynamics differ substantially between the drive cycles. While the urban delivery cycle contains many acceleration and braking maneuvers, resulting in potentially more recuperable energy, the long haul delivery cycle consists of mostly constant speed drives. \\Overall, through their vastly different characteristics, these three drive cycles should offer a good overview of possible driving scenarios.

\subsection{Electrified Trailer}
As VECTO does not natively support the simulation of recuperation axles and electrified trailers, some modifications must be made to simulate such a system. A first investigation of possible generator and gearbox combinations was done in a previous study \cite{zeller}, on the results of which our system is based.
To simulate the diesel consumption caused by the towing, the torque caused by the generator is added to the VECTO tool calculated torque as input for the engine map. The simulated trailer uses two 15 kW generators in combination with a battery of 20 kWh. The TRU is running with a constant load of 7kW, which is roughly the amount found during real-world tests by \cite{tru_consumption}.

The cycles start with a SoC of 90\% and are counted as completed when the SoC never reaches below 20\%. This threshold is set to ensure that the battery retains sufficient capacity to handle unforeseen breaks without the risk of depletion. The towing mode can be either turned on or off, resulting in the applied torque being the maximum torque of the generator at that specific speed. The recuperation mode is activated as soon as the vehicle slows down more than the driving resistances imply, i.e., active braking would need to be performed.

It’s important to note that current Electronic Braking Systems (EBS) do not allow prioritized braking with a recuperating axle. However, EBS manufacturers, such as ZF \cite{zf}, are actively developing such systems, making their existence a plausible assumption in the near future.

\subsection{Towing Optimization}
\label{opti}
In order to find the upper limit for a predictive EMS, the theoretically possible optimum for each drive cycle is calculated. For this, the additional diesel consumption caused in the truck is used as a loss function. The simulation is run repeatedly. If the SoC of the trailer falls below 20\%, the simulation will be stopped, and towing will be initiated at the most efficient time points before the SoC drops below 20\% and where towing has not yet been activated. The efficiency of towing at each time point is calculated as the towing energy in kWh divided by the additional fuel consumption in liters caused in the truck. After the towing was activated in the most efficient time points, the simulation was run again, finding possibly a new point where the SoC drops below 20\%. When the simulation finishes without a drop in SoC under 20\%, the global optimal solution for towing is found, as it is guaranteed that the minimum possible amount of towing is done and that if towing is required, it happens at the most efficient route segments. 

\section{Results}

To have a baseline for the maximum and minimum additional diesel consumption of current reactive systems, the two optimal SoC thresholds used for the bang-bang control of the towing mode were found for each drive cycle. 

This was achieved by trying out all possible combinations with a step size of 5\% SoC in between. The results are shown in Figures 4 to 6, and the best threshold values are summarized in Table \ref{table}.
\begin{figure}[h]
\centering

\includegraphics[width=0.45\textwidth]{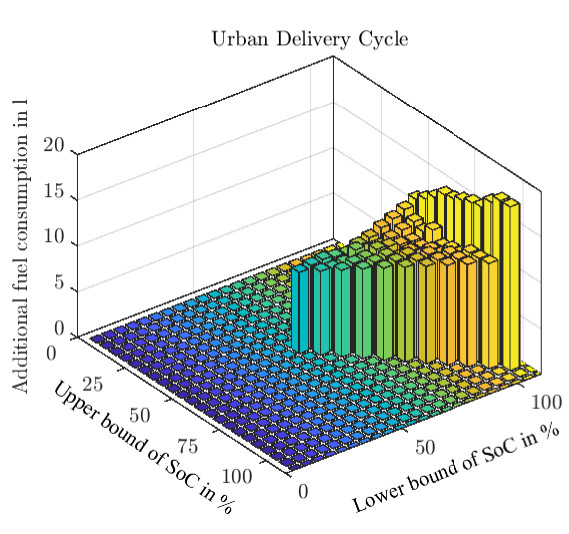}
\vspace{-3mm}

\caption{The optimal lower and upper bounds for the urban delivery cycle are 45\% and 50\%, resulting in an additional fuel consumption of 8.9l in the truck}
\vspace{-4mm}
\label{urban_bounds}

\end{figure}
\begin{figure}[h]
\centering

\includegraphics[width=0.45\textwidth]{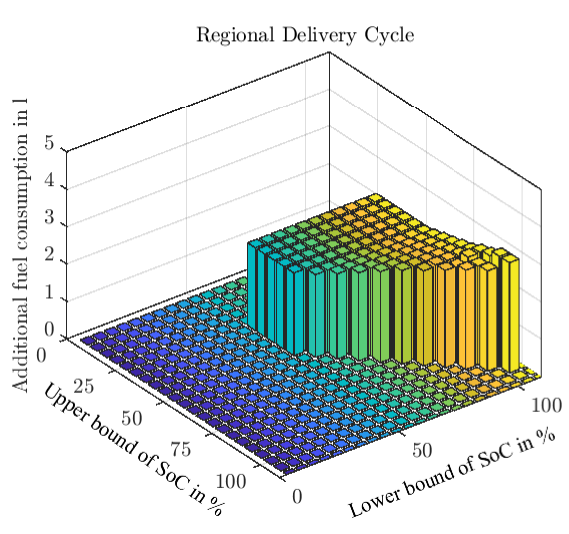}
\vspace{-3mm}

\caption{For the regional delivery cycle, the optimal lower and upper bound values are 25\% and 50\%, resulting in additional fuel consumption of 2.2l of diesel}

\label{regional_bounds}

\end{figure}
\begin{figure}[h]
\centering

\includegraphics[width=0.45\textwidth]{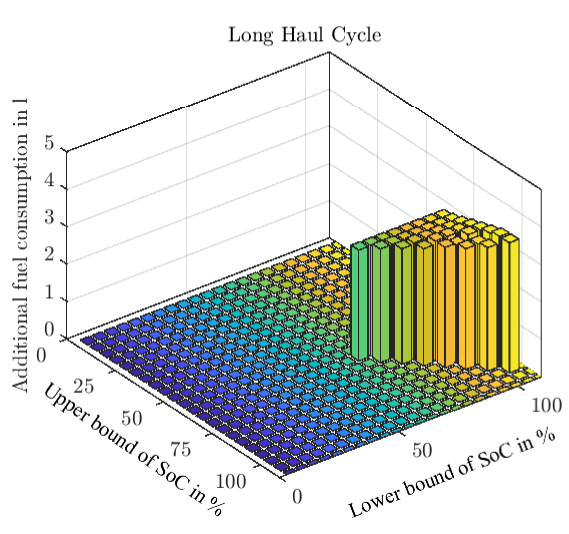}
\vspace{-3mm}

\caption{For the long haul drive cycle, the optimal lower and upper bounds are 60\% and 65\%, resulting in additional fuel consumption of 3l}
\vspace{-5mm}

\label{long_bounds}

\end{figure}

When a drive cycle cannot be completed with the current threshold values without dropping under 20\% SoC, the additional fuel consumption is set to zero in Figures 4 to 6. Overall, the most efficient threshold values are the highest possible threshold values for the lower and smallest possible values for the upper bound. However, the upper bound did not have such a significant influence on diesel consumption because it was not triggered in many of the possible combinations.

As seen in Table \ref{table} and Figures 4 to 6, the lower and upper bounds for the activation of the towing mode vary widely. This further indicates the need for a predictive energy management system, as in the final product, the bounds would be set to ensure that even the worst drive cycle would be completable with the trailer. This, however, results in further inefficiencies in all other driving scenarios.

To show the potential of a predictive energy management system, we will compare the bang-bang control algorithm typically used in today's recuperating axles with the globally optimal solution for each drive cycle. In it, towing is done as little as possible and only in the most efficient points, resulting in the minimal possible additional fuel consumption without violating the SoC boundary of 20\%.

\begin{figure}[h]
\centering

\includegraphics[width=0.45\textwidth]{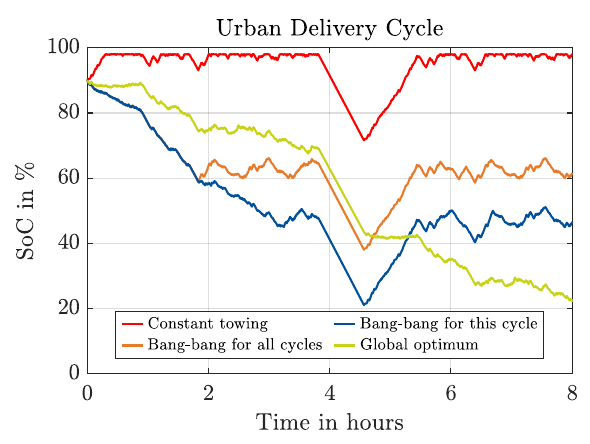}
\vspace{-3mm}

\caption{The SoC curve over the repeated urban delivery cycle shows the potential of a predictive EMS. The reactive approaches arrive with excessive energy left in the battery, while the optimization depletes the battery nearly to the limit of 20\% SoC}
\label{urban_soc}
\vspace{-2mm}
\end{figure}

The algorithm to find the global optimum is implemented as described in \ref{opti}. Compared are the traditional approach of a purely reactive bang-bang controller for the optimal values with which all drive cycles can be completed and the optimal values for the specific drive cycle, as well as a constantly towing axle as a worst-case scenario and the global optimum, which would be an upper threshold for a potential predictive EMS. The resulting SoC curves for the urban delivery cycle are shown in Figure \ref{urban_soc}. Only the optimized control depletes the battery over the drive cycle. Even with the two threshold values optimized specifically for the urban delivery cycle, the cycle ends with 45\% remaining SoC in the battery. 

This relatively high SoC at the end of the cycle is the cause of excessive towing, leading to additional costs for the operator and more $\mathrm{CO}_2$ emissions than necessary.

\begin{figure}[h]

\includegraphics[width=0.45\textwidth]{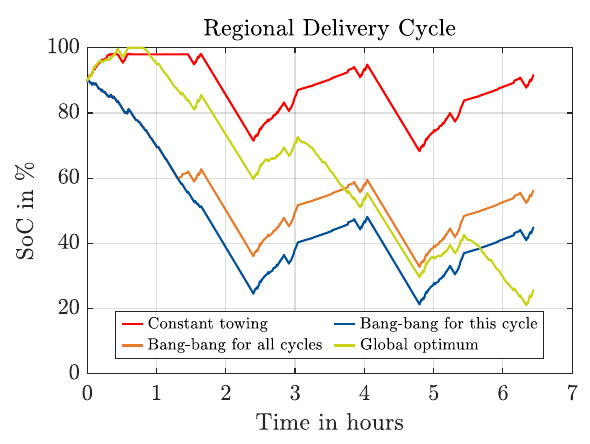}
\centering
\vspace{-3mm}

\caption{SoC curves over the regional delivery cycle. The reactive approaches arrive with over 40\% SoC }
\vspace{-2mm}
\label{regional_soc}

\end{figure}

\begin{figure}[h]
\centering

\includegraphics[width=0.45\textwidth]{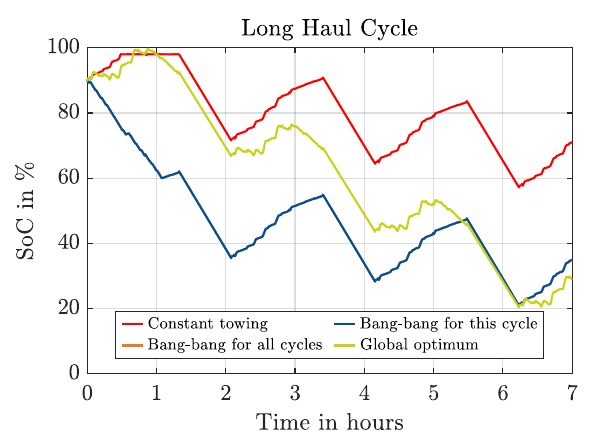}
\vspace{-3mm}

\caption{The long haul drive cycle requires the most amount of towing, leading to the smallest difference in additional fuel consumption of all three drive cycles}
\vspace{-3mm}
\label{long_soc}

\end{figure}

Similar results can be seen in the regional delivery in Figure \ref{regional_soc} and long haul drive cycles seen in Figure \ref{long_soc}, although the difference is smaller. This is caused by the fact that the system depletes its batteries, even when constantly towing in these drive cycles, reducing the potential saving of a predictive energy management system, as the higher the percentage of required towing over a drive cycle is, the lower are the possibilities to optimize the specific towing times and amounts. \\

\begin{table}[h]
\centering
\begin{tabular}{l|l|l|l}
\centering
Drive cycle    & Lower SoC & Upper SoC & Fuel \\\hline
Urban Del.    & 45\%         & 50\%         & 8.9 l             \\
Regional Del. & 25\%         & 50\%         & 2.2 l             \\
Long Haul         & 60\%         & 65\%         & 3 l              
\end{tabular}
\vspace{5mm}
\caption{Overview of optimal lower and upper bounds for the different drive cycles with the additional fuel consumption caused by the towing}
\vspace{-2mm}
\label{table}
\end{table}

In the regional delivery cycle in Figure \ref{regional_soc}, the truck arrives with more than 25\% SoC; this is caused by large amounts of recuperated energy in the final phase of the cycle with not enough time left to deplete this energy, showing the boundaries of an optimal energy management system.

For the long haul drive cycle shown in Figure \ref{long_soc}, the optimal bang-bang controller for the cycle and overall cycles are identical, as the long haul cycle required the most towing. Moreover, the large amount of towing needed leads to the drive cycle having the lowest potential savings, with 33\% of the tested drive cycles. \\The truck's additional fuel consumption results are summarized in Table \ref{main_results}.

\begin{table}[h]
\centering
\begin{tabular}{l|l|l|l}
                   & UD & RD & LH \\ \hline
Constant Towing   & 17.9 l          & 3 l                & 3.5 l      \\
General Bang-Bang & 9.8 l           & 2.4 l              & 3 l        \\
Optimal Bang-Bang & 8.9 l           & 2.2 l              & 3 l        \\
Global Optimum    & \textbf{2.5 l}           & \textbf{1.2 l  }            & \textbf{2 l    }   
\end{tabular}

\vspace{6mm}

\caption{Additional fuel consumption caused by the towing for the Urban Delivery (UD), Regional Delivery (RD), and Long Haul (LH) drive cycles. The additional fuel consumption caused by the bang-bang controlled towing can be lowered substantially, highlighting the need for a predictive EMS.}
\vspace{-2mm}
\label{main_results}
\end{table}

The urban delivery cycle has the highest additional fuel consumption during the bang-bang controlled approach and the highest saving potential compared to the global optimum. Compared to the constant towing mode, over 86\% of the additional fuel consumption can be saved, and even when compared to the optimal bang-bang controller, the global optimum requires 72\% less additional diesel fuel. This is caused by the truck driving in very different efficiency bands during the urban drive cycles, often in very inefficient operating points, giving large room for improvements by an optimization algorithm. 

On the other hand, the long haul drive cycle has the least potential for savings, as much towing is required over the drive cycle, and the truck is driving at a relatively constant speed, reducing the optimization chances. However, the results of the best bang-bang controller can still be improved by 33\% when using an optimal EMS.

\section{Conclusion and future work}

In this study, we were the first to show the potential benefits of predictive energy management systems in electrified, refrigerated trailers. We evaluated performance across these diverse driving scenarios by utilizing three distinct drive cycles from VECTO, the official simulation tool for Heavy-Duty Vehicles (HDVs) developed by the European Union. \\

Our paper identified optimal threshold values for the currently widespread approach of a bang-bang controller for each drive cycle, which were subsequently compared to the globally optimal solution.\\ For the tested cycles, the presently used approach of constantly towing or utilizing a bang-bang controller causes up to 86\% and at least 33\% higher additional fuel consumption in the truck compared to the globally optimal solution.  These findings underscore the significant potential benefits of a predictive EMS.

Looking ahead, future research will need to dive deeper into the type of data and estimations required to create an effective EMS. Key variables, such as the recuperable energy over the route and the diesel consumption at different time steps, will need to be estimated. Moreover, it is crucial to quantify the precision required for these estimates to ensure the effectiveness of the EMS. 

In conclusion, our study has laid the groundwork for further research in this field, highlighting the potential of predictive EMS in reducing fuel consumption.
\section{Acknowledgment}
The authors would like to thank the project manager ”German Aerospace Center” (DLR) for managing the project ”RekuTrAx”
(project number 01MV22018) and  the ”Federal Ministry for Economic Affairs and Climate Action of Germany” (BMWK) for the funding.

\clearpage  

\end{document}